\documentclass[conference]{IEEEtran}
\IEEEoverridecommandlockouts                              % This command is only
\usepackage{iflang}
\usepackage[utf8x]{inputenc}
\usepackage[T1]{fontenc}
\usepackage{graphicx}
\usepackage{listings}

\newcommand{\ac}[1]{#1}
\newcommand{\acsu}[1]{#1}

\title{\LARGE \bf
Topological relations in water quality monitoring
}

\author{\IEEEauthorblockN{Bruno Chaves Figueiredo}
\IEEEauthorblockA{\small INESC-ID Lisboa\\
\\
Instituto Superior Técnico \\
Universidade de Lisboa, Portugal\\
{\small 	brunocfigueiredo@tecnico.ulisboa.pt }
}
\and
\IEEEauthorblockN{Maria Alexandra Oliveira}
\IEEEauthorblockA{\small Centre for Ecology, Evolution and Environmental Changes \\
CHANGE - Institute for Global Change and Sustainability, \\
Faculdade de Ciências\\
Universidade de Lisboa, Portugal \\
{\small maoliveira@ciencias.ulisboa.pt}
}

\and
\IEEEauthorblockN{João Nuno Silva}
\IEEEauthorblockA{\small INESC-ID Lisboa\\
 \\
Instituto Superior Técnico \\
Universidade de Lisboa, Portugal\\
{\small joao.n.silva@inesc-id.pt}
}
}

\begin{document}

\maketitle
\thispagestyle{empty}
\pagestyle{empty}

%%%%%%%%%%%%%%%%%%%%%%%%%%%%%%%%%%%%%%%%%%%%%%%%%%%%%%%%%%%%%%%%%%%%%%%%%%%%%%%%
\begin{abstract}

The Alqueva Multi-Purpose Project (EFMA) is a massive abduction and storage infrastructure system in the Alentejo, which has a water quality monitoring network with almost thousands of water quality stations distributed across three subsystems: Alqueva, Pedrogão, and Ardila. Identification of pollution sources in complex infrastructure systems, such as the EFMA, requires recognition of water flow direction and delimitation of areas being drained to specific sampling points. The transfer channels in the EFMA infrastructure artificially connect several water bodies that do not share drainage basins, which further complicates the interpretation of water quality data because the water does not flow exclusively downstream and is not restricted to specific basins.

The existing user-friendly GIS tools do not facilitate the exploration and visualisation of water quality data in spatial-temporal dimensions, such as defining temporal relationships between monitoring campaigns, nor do they allow the establishment of topological and hydrological relationships between different sampling points. 

This thesis work proposes a framework capable of aggregating many types of information in a GIS environment, visualising large water quality-related datasets and, a graph data model to integrate and relate water quality between monitoring stations and land use. The graph model allows to exploit the relationship between water quality in a watercourse and reservoirs associated with infrastructures.

The graph data model and the developed framework demonstrated encouraging results and has proven to be preferred when compared to relational databases.

\end{abstract}

%%%%%%%%%%%%%%%%%%%%%%%%%%%%%%%%%%%%%%%%%%%%%%%%%%%%%%%%%%%%%%%%%%%%%%%%%%%%%%%%
\section{Introduction}

The European Union, established water policies such as the  Water Framework Directive (WFD), which focused on the evaluation of water bodies in order to improve, protect, and prevent further deterioration of the water quality in European Union (EU) \cite{eu:water-directive}.

In Portugal, sampling in primarily done in large rivers and reservoirs for domestic water supply, with water samples being taken at monthly intervals and analysed for general chemical and physical variables, organic pollution indicators, nutrients and some heavy metals \cite{Kristensen1996}.
The number of parameters to be monitored is continuously increasing, and are constantly modified and refined in response to the expanding uses of water, as well as the development of analytical capabilities for measuring more substances at ever-lower concentrations \cite{Kristensen1996}. 

The Alqueva Multi-Purpose Project (\ac{EFMA}) is located in the Alentejo region of Portugal and consists of water abduction and storage infrastructures, as well as infrastructures to improve irrigation areas \cite{Coimbra}. 
The EFMA includes several reservoirs, small-medium size dams, and transfer channels that feed the Alqueva dam, a major dam in the Guadiana river \cite{Bettencourt2009}.

The water quality monitoring system of EFMA is a network with hundreds of water quality stations mainly operated and maintained by two entities, Empresa de Desenvolvimento e Infra estruturas do Alqueva, S.A. (\ac{EDIA}) and Portuguese Environment Agency (\ac{APA} /\acsu{ARH})\label{acro:ARH-ALENTEJO} \cite{Ruivo2012,inag:redes}.

  \begin{figure}[thpb]
      \centering
      \includegraphics[scale=0.45]{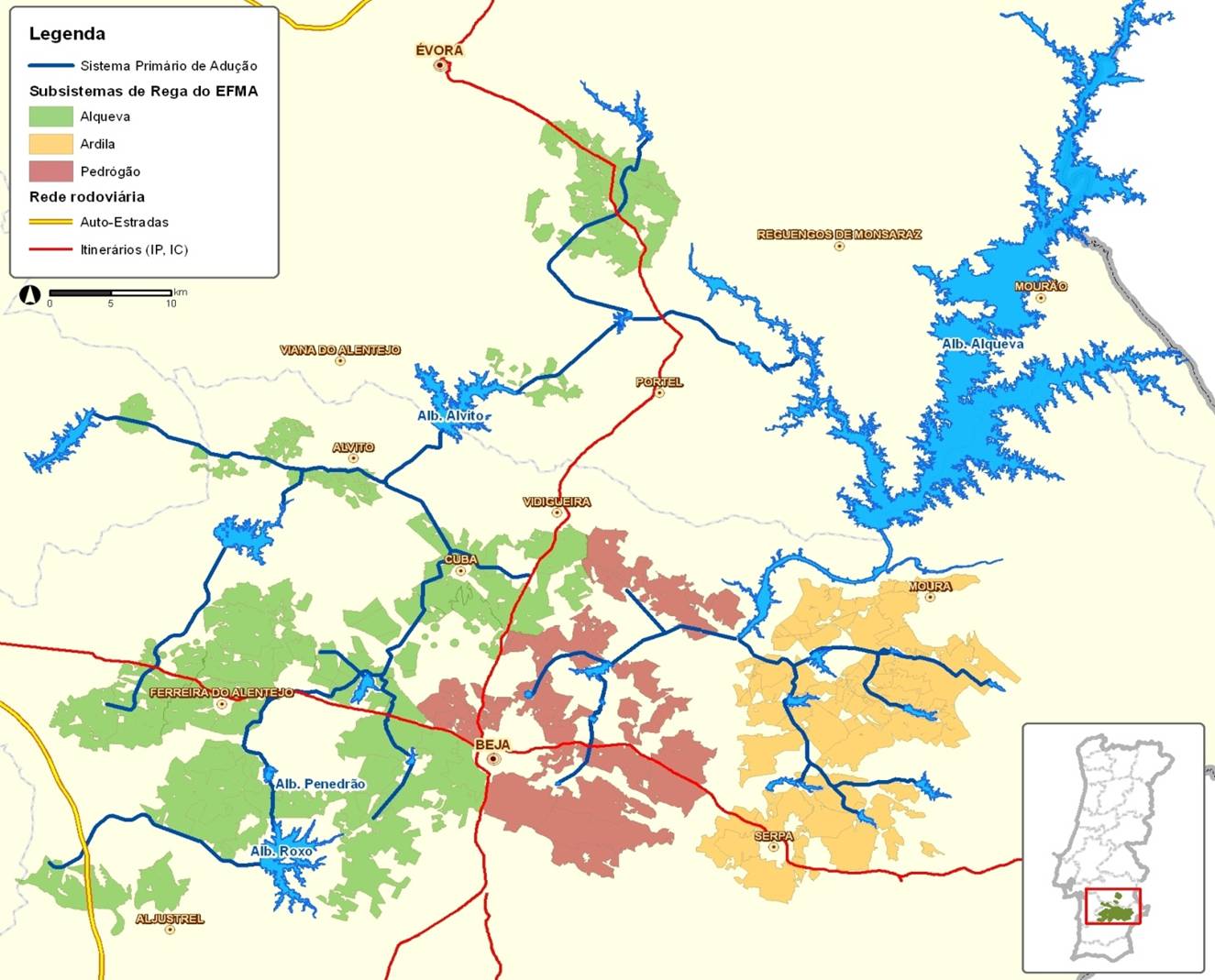}
      \caption{EFMA region, Alentejo.}
      \label{sasa}
   \end{figure}

The \ac{SNIRH} database is a public database available via web browser, that aggregates most of the available water quality data from Portugal, including the EFMA region, also storing data from old water monitoring programmes, such as  Estudo dos Recursos Hídricos Subterraneos do Alentejo (\ac{ERSHA}).

The EDIA monitoring bulletins and SNIRH database are two independent data sources and follow their own data schema. Furthermore, both the APA and EDIA institutions have stored data of field monitoring campaigns in tabular databases with inconsistent data schemas.
Currently, it is challenging to create a database with different types of complementary information, such as hydrometry (e.g, precipitation amount and water levels) or geological and land use data that is typically represented spatially explicit information, such as vectors, including the ESRI Shapefiles, or rasters.

The monitoring stations can evaluate multiple parameters and have various purposes \cite{Morais2007}. The water quality is frequently evaluated in several locations and, in dams and reservoirs or large rivers, at several depths. 
In addition, measurements include physical and chemical parameters, collected \textit{in situ} or sent to certified laboratories for chemical analysis \cite{edia:relcontas-2020}.
As a result, a large amount of 4D data (2D parameters, space, and time) that, due to the complexity associated with its representation, are not fully utilised and explored, for example, in the detection of frequent water pollution sources.
When working with 4-dimensional data, both temporal and spatial correlations must be considered, which complicates the data analysis and visualisation process significantly. % solution that can incorporate and  visualize efficiently 4-dimentional water quality  data and  GIS features

Current tools have limited capability in handling more complex data and are frequently unable of adapting to new methods, standards, and criteria used on water quality monitoring.
The increasing parameters and number of sampled data force researchers and managers to find customised scalable solutions that need to be flexible to work with new types of data and data relationships, and that can evolve the advancing progress of technology and scientific research work.

The creation of a framework that can detect water relationships between sites on a river's branches and correlate them with water quality samples would be a significant tool for researchers to quickly monitor contaminant circulation over the river network and water distribution systems. Also, the developed solution offers a tool to migrate these spreadsheets to a database, which would reduce data redundancy and provide data abstraction and advanced querying.

%%%%%%%%%%%%%%%%%%%%%%%%%%%%
%%%%%%%%%%%%%%%%%%%%%%%%%%%%
%%%%%%%%%%%%%%%%%%%%%%%%%%%%
%%%%%%%%%%%%%%%%%%%%%%%%%%%%
\section{Related Work}

\subsection{Water pollution}
%%%% Discussion water pollution
Water pollution is defined as any physical, chemical, or biological change in water quality that has a detrimental effect on living creatures in the ecosystem or renders a water resource unfit for one or more of its intended uses \cite{bookWHO}
Currently, commercial software is uncapable to address all these issues and the development of a customised tool that integrates and stores this large amount and diversity of the data is needed for this kind of data processing and analysis.
The chemical usage in agriculture can affect directly the water quality over the terrain drainage which reach surface waters such as streams, ponds and may compromise the enriched aquatic habitats important for aquatic species. The phosphate, nitrogen, nitrates, and organics are the main chemical substances that associated to livestock farming \cite{apa:aia:vcastelo,eia:inf12} that promotes eutrophication. Eutrophication is the process of accelerated nutrient enrichment of a water body \cite{phdLindim}, which contributes to enhanced aquatic plant growth that negatively influences water quality and, in particular, dissolved oxygen availability \cite{Boyd2020}.
%%%%

\subsection{Alqueva Multi-Purpose Project}

The Environmental Management Programme (EMP) of EDIA was approved by Joint Ordinance 1050/2005 of 6 December of the Ministry of the Environment, Territorial Planning and Regional Development and Ministry of Agriculture, Rural Development and Fisheries, and provides the promotion and coordination of the design and implementation of environmental monitoring programmes for different aspects of the EFMA.
The EDIA is responsible for monitoring environmental factors in the EFMA, from status of surface and underground bodies of water, to soils, fauna and flora \cite{edia:social-policy}, and is responsible for implementing and researching new climatological and hydrometric monitoring stations \cite{edia:wwd-monitoring}.
The WFD, which has been transposed into the national legal system via the Water Law, establishes the framework for a sustainable water management, highlighting the main environmental goal of achieving a “Good” status for all the surface water and ground water bodies \cite{apa:en-about}.

Additional water quality assessments are undertaken by the Portuguese Environment Agency (APA) and made available to the public through the National Water Resources Information System (SNIRH), which is a database that compiles data from a monitoring network that covers all the regions of Portugal, including the EFMA region.

The SNIRH processes, validates and publishes all the information collected in the APA monitoring networks and those of other bodies (e.g., local and regional, including in the Azores and Madeira archipelagos).
The SNIRH helps achieve the following objectives: assessment of water availability; characterisation of water resources; assessment of the water quality trends; among others \cite{apa:rea-water}.

In addition to the EDIA and APA water quality monitoring database, groundwater quality data from early environmental impact studies done by ERSHA (Estudo dos Recursos Hídricos Subterrâneos do Alentejo) over the 1997–2001 period are also available \cite{apa:snirh}.
At the time, the Portuguese Water Institute (INAG) \cite{AntonioReisRosadoParalta2009}, part of APA since 2012 \cite{inag:erhsa}, was responsible for the study of the aquifer systems.

\subsection{Geographical Information System}
%%%% Discussão %%%% Geographical Information System
%The disadvantage of the "WaterMarque" approach is that developing and maintaining a graphical user interface is a complex task that takes up the bulk of the system development time, which limits the time available for developing new analytical and presentation techniques. The ideal approach is to work with a combination of two methods, retaining complex procedures in a batch-mode processing environment and building generic procedures into an interactive system.
%%%%

 %%%% Discussão %%%% Geographical Information System
In the geospatial analysis process, maps are extremely important\cite{Peuralahti2014} because they are a type of graphical user interface (GUI) with a geospatial dimension that provide a GIS direct and interactive interface \cite{Peuralahti2014}. They also help evaluate the intermediate analysis results and present the final results \cite{kraak2010cartography}.
%%%%

GIS has become a highly effective tool in processing and representing environmental data. GIS is capable for handling large amounts of spatial data in order to model non-point source pollution problems \cite{Tim1994} and combines native raster and native vector data that are commonly used in environmental assessments \cite{articleWade}.

% \ac{GIS} has become a particularly powerful and vital tool in processing and representing environmental data.

% %%%%% Disucssão GIS applied to Water Quality
% GIS approaches are useful for handling large amounts of spatial data in order to model non-point source pollution problems\cite{Tim1994}. 

% GIS-based measurements that combine native raster and native vector data are commonly used in environmental assessments \cite{articleWade}. 

% GIS is a highly effective tool for water quality modelling because it combines information from several sources to form a model \cite{Tim1994}. 
\subsubsection{GIS applied to Water Quality}
%%%%% Disucssão GIS applied to Water Quality
Most hydrological analyses performed using GIS frameworks are static, with data processing and representation being performed separately or in combination, but not as an integrated process: the spatial analysis is conducted using GIS, and the time-series analysis is mostly done using spreadsheets or statistical software. Neither approach considers the source-sink relationships between sampling points. Alternatively, some authors choose to highlight the changes in water pollution observed throughout time, which is a classical way to evaluate water pollution and does not require the use of GIS. This way, time-series analysis is performed and the tabular data that do not include either location (and hence, distance), or topological relationships, are statistically analysed \cite{Morais2007,Palmaarticle,Perez2009,McLeod1991,MATTIKALLI1996149}%(McLeod et al., 1991; Morais et al., 2007; Palma et al., 2010; Perez et al., 2010; Mattikalli, 1996).

Geospatial data are not easy to store in a standard RDBMS. As a result, spatial extensions have been developed, such as PostGIS, a PostgreSQL add-on that provides standard spatial data types and spatial queries.\cite{Jolma2006}. However, graph database technology has been increasingly used and has been found to handle graph-like spatio-temporal data much more effectively \cite{Effendi2020}. In the previous section, it was highlighted on some studies that graph databases outperformed PostGIS/PostgreSQL on interconnected spatio-temporal data.

\subsection{Graph Databases}
% % %%%% Graph Databases 2.4
The GIS databases are the underlying support for web GIS applications \cite{Peuralahti2014}. The GIS data storage should hold a collection of geographic datasets of different types, such a basic vector data (e.g. points, lines and polygons) and raster data such as satellite and aerial images \cite{Peuralahti2014}. 

There are important constraints on its use for the representation of water pollution: water is always moving, and the position of a water particle or a water contaminant varies geographically and temporally \cite{Allafta2020,Lintern2017,Hugo2005}, following the hydrological cycle.

 %%%% Graph Database Performance
The flexibility of a graph model permits the addition of new nodes and relationships without compromising the existing network and storage system and without expensively migrating the data \cite{Sasaki}.

%%% Discussão Graph Database Performance
The existing approaches using relational databases have encountered performance limitations, both in storing and searching for information, when processing large amounts of data. In this domain, the datasets are fairly large and describe a wide range of connections and entities \cite{Gong2018}. %Figure ~\ref{sasa} illustrates the positioning of the graph databases versus relational and other NoSQL.
%%%%%

The use aggregate stores for interrelated data results in a disjointed development experience since it must have to add a lot of code to fill the limitations of aggregate store storage principles. Furthermore, as the number of hops (query complexity) increases, aggregate stores slow down significantly \cite{Sasaki}.

%   \begin{figure}[thpb]
%      \centering
%      \includegraphics[scale=0.13]{dbsprectrum.png}
%      \caption{The spectrum of databases for discrete versus connected data \cite{Sasaki}}
%      \label{sasa}
%   \end{figure}

\subsubsection{Graph Data-driven approaches for River management}

Graph databases open up new possibilities as to how we can efficiently access and interconnect data, while recalling the performance of the graph theory on complex querying and data searching. This way of storing information brings new approaches to current problems that were limited with other database solutions. Graphs appear to be capable of advancing current studies on water quality assessment or river network modelling in hydrology and water management \cite{2020EGUGA..2217318F,articleTorres}.
A recent study demonstrated that graphs can be used to model geospatial data in hydrology (e.g. watersheds or drainage networks) \cite{Jaudet2017}. A watersheds are usually connected in a drainage network with confluences, start and end points connected by drainage stretches (network edges)\cite{Jaudet2017}.

\begin{figure}[thpb]
      \centering
      \includegraphics[scale=0.2]{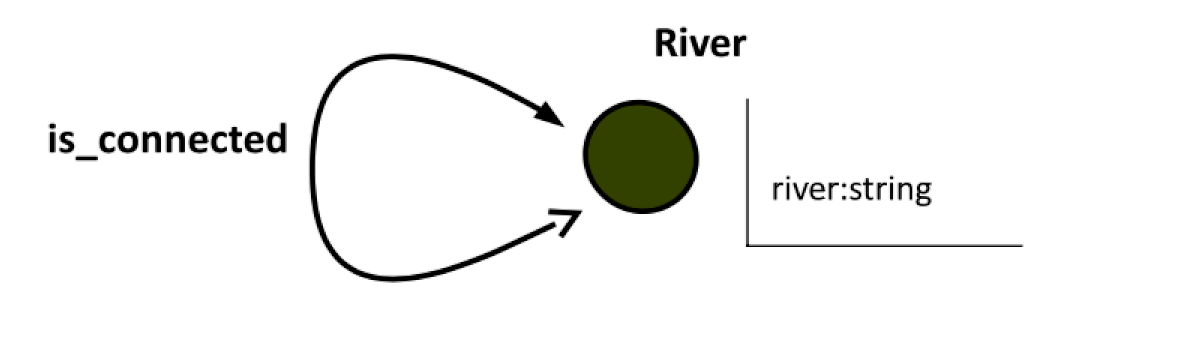}
      \caption{GVRiver graph view schema. Adapted from \cite{Jaudet2017}}
      \label{fig:riverg}
   \end{figure}

A property graph data model, which is based on labelled-property graphs, was proposed to facilitate the understanding of drainage networks \cite{Jaudet2017}. The study presents a use case for modelling a drainage network and watershed hierarchies where drainage points, rivers and watersheds are represented as nodes of a graph, containing property information, such as the ID and type, river name, and watershed ID code, respectively. The relationships between drainage points represent the water stretches. In the presented data model, a river has an associated drainage network and can be connected to other rivers through a drainage point (Figure ~\ref{fig:riverg}). Additionally, the watersheds have associated water stretches and can be part of other watersheds \cite{Jaudet2017}.

According to that study, river systems and other types of transportation networks can be modelled as graphs and implemented using graph databases, on which queries are more easily expressed using high-level graph query languages, such as Cypher. Its findings also show that queries involving path computation run faster on graph databases, in general, without using advanced SQL query tuning, due to the fact that their underlying data structures are optimised for a fast path traversal. A relational representation, on the other hand, requires the use of recursive queries to compute the transitive closure of a graph, which reduces query efficiency because the relational representation does not adequately capture the graph structure. Aggregate, path, and spatial queries were among the five types of queries investigated. Only three of the 16 queries performed better in the relational version. The results were obtained using Neo4j library algorithms rather than the built-in transitive closure computation of Cypher. Long path traversals, such as those required in this problem, were clearly not handled appropriately when using the relational model, as they require the multiple self-joins of the table containing the relationships between the river segments \cite{Bollen2021}.
%%%%%%%%%%%%%%%%%%%%%%%%%%%%%%%%
%%%%%%%%%%%%%%%%%%%%%%%%%%%%%%%%
\section{Data Modelling}
%The established relationships between different features and directions of water flow will allow the evaluation of water and contaminant paths.
%Classically, water pollution has been attributed to specific landuses/sources by considering the drainage area (watershed) located up-flow from the point of detection. This has allowed to identify pollution sources and also to define local measures that mitigate further contamination. In a complex system such as the case study addressed herein this simpler approach no longer works due to artificial channels and pumps that allow water transfers within and between  different reservoirs/dams and systems (Pedrogão, Ardila and Alqueva). Meaning that a pollutant can be originated in a sub-watershed and contaminate other streams that are not typographically connected.  %This approach will contribute to a better management of these types of complex systems.
The information was available in 4 different file types: spreadsheets (e.g., water quality data), ERSI shapefies (e.g., watersheds, soil and land use, EFMA infrastructures and reservoirs), TIFF format (e.g., \ac{DEM}) and PDF, which contains information about of the EFMA water distribution system.

\subsection{Internal Representation}

The new data model aim to integrate and relate water quality between monitoring stations and land use, as well as exploit the relationship between water quality in a watercourse and reservoirs associated with infrastructures. By modelling the EFMA system as a graph data model, it is possible to connect all data elements through their relationships and unlock new analysis methods given the graph model characteristics, such as relationships between infrastructures to search for paths to the origin of contamination via specific chemical parameters or to identify the water sub-systems that will be affected by the contamination via network paths.

\begin{figure}[h]
\centering
\includegraphics[scale=0.29]{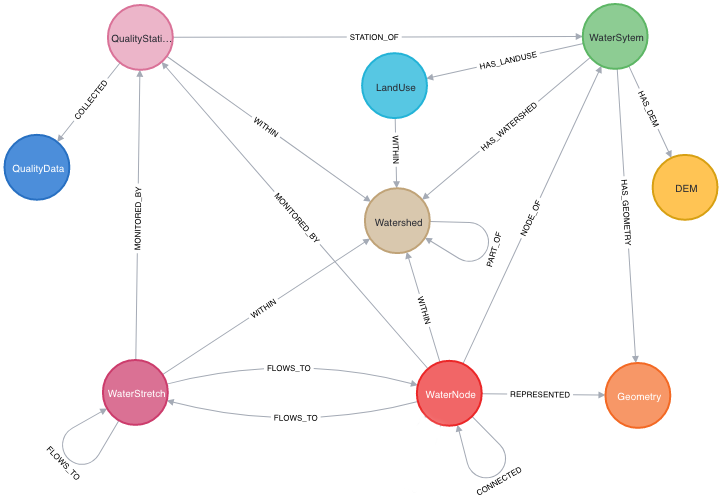}
\caption{Metagraph data schema} 
\label{fig:metagraph}
\end{figure}

 The figure ~\ref{fig:metagraph} presents the proposed data model for the EFMA water quality system which consists into a directed graph with node entities that models the river water stretch, the water quality monitoring network, the infrastructure network, and water basin characteristics (e.g., sub-watersheds and land use), as well as the relationships between them, in a spatial referenced context.

\subsubsection{Nodes}

The proposed data model has the following list of entities: $WaterSystem$, $WaterNode$, $QualityStation$, $QualityData$, $WaterStretch$, $Watershed$, $LandUse$, $Geometry$ and $DEM$.

%The EFMA water subsystem diagrams are modelled with nodes representing the infrastructures and water systems. 

The $WaterSystem$ nodes on the data model represents the water system. Associated to $WaterSystem$ are the infrastructures, a water monitoring network and GIS data, such as raster with elevation data (e.g. DEM).

The $WaterNode$ represents the physical elements associated to a water system, the infrastructures.The $QualityStation$ nodes represent the water quality monitoring stations. The data quality data collected by a water quality station is represented by the $QualityData$ node. The GIS data integrated to this thesis project is divided into vector and raster data. Vector data is composed by $Geometry$, $Watershed$, and $LandUse$ nodes, while raster data consist of $DEM$ nodes. The $WaterStretch$ node, which has the id and geometry properties, represents a stretch of the drainage network that flows to and from infrastructures.

\subsubsection{Relationships}

The Node entities can be interrelated by using the following relationships: $\mathord{:}CONNECTED$, $\mathord{:}MONITORED\_BY$, $\mathord{:}STATION\_OF$, $\mathord{:}COLLECTED$, $\mathord{:}FLOWS\_TO$, $\mathord{:}WITHIN$, $\mathord{:}HAS\_LANDUSE$, $\mathord{:}HAS\_WATERSHED$, $\mathord{:}HAS\_GEOMETRY$, $\mathord{:}REPRESENTED$, and $\mathord{:}PART\_OF$.

The $\mathord{:}CONNECTED$ relationship specifies the water connection between $WaterNodes$, which represent the artificial water transfers that connects the infrastructures. The attribution of vector data to a $WaterNode$ is made using $\mathord{:}REPRESENTED$. The relation $\mathord{:}FLOWS\_TO$ is established  between $WaterStretch$ A and B nodes if water can flow from segment A to segment B. The relation $\mathord{:}MONITORED\_BY$ establishes the relation between the $WaterStretch$ (or $WaterNodes$) and the water quality station.  The $\mathord{:}HAS\_LANDUSE$, $\mathord{:}HAS\_WATERSHED$ and $\mathord{:}HAS\_GEOMETRY$ relationships indexes the GIS data associated with a $\mathord{:}WaterSystem$, such as $LandUse$, $Geometry$, and $Watershed$, respectively. The $\mathord{:}WITHIN$ indicates that a $QualityStation$, $WaterNode$, $WaterStretch$ or $LandUse$ is located with a specific watershed.  $\mathord{:}PART\_OF$ is used to indicate sub-watershed that are within a watershed.

\section{Implementation}

 The proposed architecture is described and essential components are explained in this section. Furthermore, the implementation steps of the Framework are also addressed.

\subsection{Framework Functionalities}
The main goal of the thesis work, as stated in the first chapter, is to develop a framework capable of aggregating all different information in a GIS environment, visualising and analysing large water quality related datasets, and exploring interactively the topological relationships between data. The framework is a web application designed to aid in the management of water distribution networks and assist the water quality assessment of a certain region. The framework must support the following functionalities:

\begin{itemize}
\item Input file processing and data conversion for database storage: A file upload form for processing and populating information from files mentioned before into the database. The upload form accepts CSV files, such as water nodes, water nodes links, water quality stations, water quality data, as well as the geospatial data in GeoJSON or ESRI shapefile formats, such as watersheds and land use.

\item Interactive map for visual geodata exploration: Simple user interface for creating and editing database information. A simple web map for visualising the water distribution network and exploring geospatial-related information (e.g., water nodes, water quality stations, and watersheds). The web map interface also provides methods to create and delete data from the database.

\item Data filters and big-data visualisation toolbox for water quality data analysis: A console interface for filtering specific parameters from water quality data and visualise data into bar, time series and bubble charts.  
% : Data visualisation of water quality associated with a river basin. 

\item Export tool for water quality analysis results: Export the water quality data results from console to CSV files.

\item Path-finding discovery tool between points in the water distribution system network: Network routing of the water distribution graph and displays the routing results on the web map. It features options such as finding the water sources of a water node and discovering the path between a water node and the end points of the network.
%\item Tool for identifying spatial relationships between water quality stations, drainage networks, watersheds, and land use data.
\end{itemize}

To gather the information that was not available in the sourced files, such as the water stretches of drainage networks and the association between water stretches and water quality stations, I developed a toolbox to compute and estimate sub-watersheds, drainage networks, and some geometric manipulations for geospatial features, such as the intersection between polygons and points. This tool will be described in the next section.

\subsection{Framework Architecture}

\begin{figure}[h]
\centering
\includegraphics[scale=0.13]{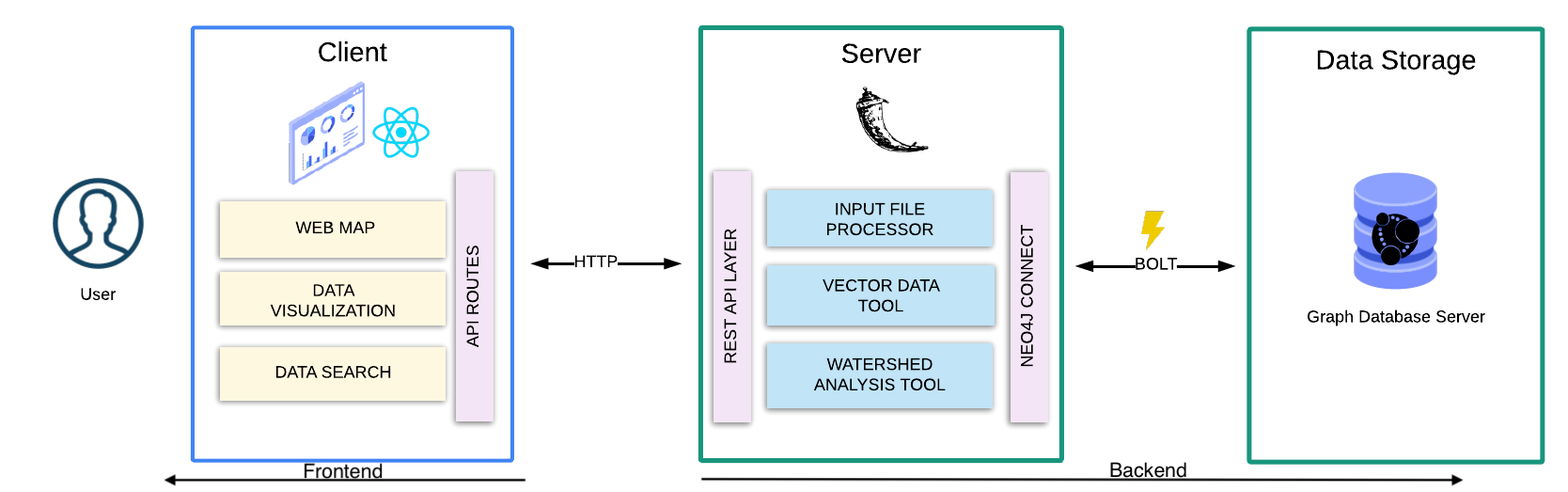}
\caption{Client-Server Architecture.} 
\label{fig:architecture}
\end{figure}

The architecture of the web application consists of a 3 tier client-server architecture with: client, web server and database server. This architecture allows to isolate the development and maintenance of each tier, and allows to run the tiers separately. Also, the mid-layer adds another security level to the database. Figure ~\ref{fig:architecture} presents the architecture proposal for the web application.

The client-server architecture is very popular for common web applications and, as mentioned in chapter 2, is also used for many web GIS applications. To run the web application on old machines or low-performance machines, the GIS processing tasks occur at the server, while the client is only used to show results and user interactions (i.e., definition of a thin architecture).  

The client is an application that runs on the web browser. The client provides the user-interface for data visualisation and exploration. The client comprises 3 modules: a web map, search data, and data exploration. The web map displays all geographic data accessible in the database, and the search data module is where the user searches and requests data from the database. The data exploration module is the interface to analyse, filter and plot the water quality data. The client communicates with the back-end over the HTTP protocol.

The back-end consists of an API server that links the user and the database. The server makes the resources available through REST API to the client. 
The REST API provides a uniform interface and client-server separation that makes the development of client and server more independent of each other.  

The server has all GIS-related computational and input file parsing processes described by the modules: input file processor, vector data tools and watershed analysis tool. The input file processor is the module that parses and converts the files, if needed, before populating the database. The vector data tool consists of a set of vector data methods (e.g., intersection and contained verification methods). This module is used to verify relationships between data in a GIS context, such as to verify if a watershed is within another watershed, or to detect if a water quality station intersects a water line. All drainage network and watershed calculations are performed in the drainage analysis module. 

The server communicates with the database via bolt protocol. Bolt is a binary protocol that is more compact and has better overall performance than HTTP due to the binary serialisation format for the requests and responses, making them smaller and faster to send large chunks of data. At the last tier, the database server stores all information in a graph data model.

\section{Framework Implementation}

\subsection{Frontend}

The Frontend was developed in Javascript with the ReactJS library. Javascript is the most popular programming language for frontend and is the default runtime of the web browsers. 
 The application is made up of five main UI components organised in a grid layout, that can be resized and movable. The main UI components are: Console, Map Search, Search, Navbar and Map Settings. 

The Map Search component provides the web map interface to explore the geospatial data and relationships. The Search component is for searching the infrastructures and centring the web map on the finding results. The Map Settings is the interface for controlling the data displayed on the web map, and the Navbar is used to control the application background colour and manage the other components at the layout grid, such as adding a new component to the grid. The console component is for uploading the input files, visualising the selected quality stations, and plotting  water quality data from selected quality stations. 
\subsection{Backend}

The backend was developed in Python, a very popular language along with JavaScript, and has many mature libraries for creating a web server and handling geospatial data. The web server was implemented using the Flask \cite{flask}, a lightweight python-based web framework that does not include database abstraction layer, form validation, and other features that are common to other frameworks, such as Django \cite{django}. Flask is simple to implement since it does not enforce any dependencies or project layout, and supports extensions to add functionalities to the server. In this implementation, some extensions were used:
\begin{itemize}
\item \textbf{Flask-Caching}: adds caching support for various backends to any Flask application.
\item \textbf{Flask-JWT-Extended}: adds support for using JSON Web Tokens (JWT) to Flask for protecting routes and many features built in to make working with JWT easier.
\item \textbf{Flask-Executor}: easy to use wrapper for asynchronously executing callables.
\end{itemize}

\subsubsection{API}

The web server uses RESTful APIs to communicate with the frontend application. the available APIs can be categorised into Authentication API, File API, Queries API and Resource API. 

The Authentication API is used to authenticate the client and granting access to all other APIs. The API uses JWT-based authentication. The File API includes several endpoints for uploading files to the database, including specific file upload endpoints and a main endpoint that auto-identifies the file type and redirects the file to the appropriate endpoint for processing. 

The Queries API has \textbf{42} endpoints for querying almost all types and forms of data from the database. All endpoints are cached to minimise the number of requests to the database. The resources API has endpoints for GIS computational tasks, such as calculating the sub-watersheds or GIS data manipulations. Because some computations can take long execution times, the endpoints use parallel and asynchronous tasks to make the result available later.

\subsubsection{Database}
The graph database solution chosen to this project was Neo4j \cite{neo4j:cypher}. Neo4j is a native graph technology that has substantial support, such as extensive documentation and reference code, and offers spatial functions for exploring geospatial data stored in the database. It is also simple to setup and offers a free online sandbox for testing queries. Neo4j has its own query language called Cypher.

\section{Evaluation and Validation}

In this section, we analyse the potential of the implemented framework to solve the challenges described before. We evaluate the implementation using a graph database, such as the effectiveness of the data model introduced in section ~\ref{sec:internal} and highlight the differences between the queries implemented in Cypher with a solution in SQL. Furthermore, we test the essential functionalities offered by the Framework, which include both frontend and backend performance and the ability to visualise time-series water quality data.

\subsection{EFMA Dataset}
The dataset used to evaluate the platform implementation is composed of files that resulted from pre-processing using the procedures described in the data preparation section: Quality stations CSV file, CSV files with the water nodes of Alqueva, Pedrógão and Ardila, GeoJSON geometries of the water nodes, GeoJSON file with the hydrographic basins, GeoJSON file with the land-use and the several processed water quality files.

The EFMA data produced a graph with \textbf{44887} nodes and \textbf{44285} edges. More accurately, there are 115 water nodes, 795 water quality stations, 43892 water quality data, 39 geometries, 23 watersheds, 22 land use zones and 1 DEM. Furthermore, the stored DEM generated 1862 nodes (i.e., water stretches) and 1922 edges. The resulting graph database has a storage size of approximately \textbf{0.2GB}.

\subsection{File Upload Performance}
\label{sec:upload}
We benchmark the performance of water nodes and water quality data uploads using the Framework, analysing the behaviour of information size versus processing time. For comparison, we measured the upload and processing times of other files. File processing time includes the following components: web application processing time, communication time between the three layers, server processing time, and database populating time.

The upload was tested using the API server endpoints dedicated to processing each file type. The test was carried out on a single machine equipped with a CPU 1,8 GHz Intel Core i5 and 8 GB 1600 MHz DDR3 RAM memory.  the upload times were measured at the web application using the developer tools of the Opera browser. In the water quality files test, the larger files take substantially longer to process, which may be explained by the files being processed on the web server before being uploaded to the database.

\begin{figure}[h]
\centering
\includegraphics[width=0.35\textwidth]{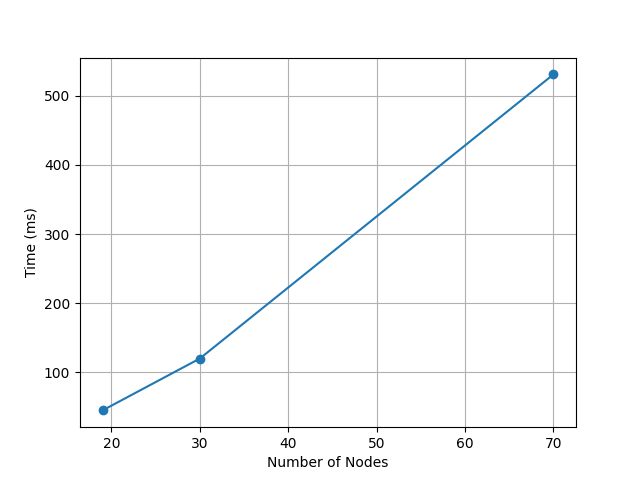}
\caption{Upload time versus number of nodes.} 
\label{fig:waternodesupload}
\end{figure}

The upload time with the increase of the number of water nodes is shown in Figure ~\ref{fig:waternodesupload}. Although the water node files show acceptable processing times, they only represent a practical case of uploading small files. The increase in time with much larger files, on the other hand, cannot be predicted. 
The GeoJSON files upload times are: 9s, 0.3s and 2s for land use, watersheds and geometries, respectively. We also evaluated the data in ESRI shapefile format, which took 16s for land use (6700kB), 0.55s for watershed (2200kB), and 5.1s for geometry (129kB).

\subsection{Database Query Analysis}
The potential and capabilities of the Cypher queries developed in this project are explored in this chapter. We will go over four main queries (i.e, namely Q1, Q2, Q3 and Q4) that will allow the user to explore the EFMA network water relationships and discover relationships between water quality data.
The following queries were run on a single machine equipped with a 1.8 GHz Intel Core i5 CPU and 8 GB 1600 MHz DDR3 RAM memory. The Neo4j database console was used to measure all of the time records.

\subsubsection{Q1: Finding water source of a water node} % procura de caminho source

The query  is used to locate the water source of a given point in the EFMA water transfers network. The query is simple and descriptive, which is a Cypher characteristic that allows developers to write explicit relationship patterns in queries. Neo4j provides a simple path finding procedure that consists of assigning a variable to the relationship pattern (i.e. path variable in ~\ref{lst:findnode}). The query runtime for a graph with 115 water nodes and 113 edges, the longest path of which has 21 edges, was 3ms. The runtime remained consistent across multiple runs.

%\begin{lstlisting}[caption={Find water source of a given water node.},frame=tlrb, label={lst:findnode},  language=SQL]{Name}
%MATCH path=(source:WaterNode)-
%[rel:CONNECTED*]-
%>(end:WaterNode{id:$id})
%WHERE NOT (()-[:CONNECTED]-
%>(source))
%RETURN source
%\end{lstlisting}

%%%%%%%%%%%%%%%%%%%%%%%%%%%%
\subsubsection{Q2: Finding full water paths containing a given water node} % procura de caminho em que o no esta contido 

The query identifies find the path between water source and end locations for a given infrastructures. This query is used to identify the upstream and downstream infrastructures of given infrastructure.

%\begin{lstlisting}[caption={Identify the upstream and downstream infrastructures.},frame=tlrb, label={lst:findpathnode}, language=SQL]{Name}
%MATCH p=(source:WaterNode)-
%[:CONNECTED*]->(:WaterNode{id: 
%$id})-[:CONNECTED*]->(end:WaterNode)
%WHERE NOT (()-[:CONNECTED]-
%>(source)) AND 
%NOT ((end)-[:CONNECTED]->())
%RETURN collect(p) as path, 
%count(p) as npaths
%\end{lstlisting}

The query is similar to the previous one, but it searches for the complete path pattern (i.e., backward and forward) and returns a list of paths. When compared to an SQL implementation, the SQL version would be more complex in determining the path between nodes with nodes and edges stored in tables. Recursive queries with multiple joins between the two table entities may be required to find a path between two nodes. The query took 6ms to return the longest path on the EFMA distribution network in the console (i.e., path with 21 edges). The query was run multiple times, and the time consistency was maintained.

\subsubsection{Q3: Finding all quality stations in a downstream watercourse} % % procura de estacoes ao longo do stretch em sentido downstream
The query is used to determine the entire path of a watercourse and collect all monitoring water quality stations along that path. The query begins by locating the full path watercourse that contains a specific water stretch, and then searches for monitoring water quality stations associated with the discovered water stretches.

%\begin{lstlisting}[caption={All water quality stations located downstream watercourse.},frame=tlrb, label={lst:downstream}, language=SQL]{Name}
%MATCH p=(source:WaterStretch)-[:FLOWS_TO*]-
%>(:WaterStretch{id:$id})-[:FLOWS_TO*]-
%>(end:WaterStretch)
%WHERE NOT (()-[:FLOWS_TO]-
%>(source)) AND 
%NOT ((end)-[:FLOWS_TO]->())
%UNWIND nodes(p) as n
%MATCH (qs:QualityStation)<
%-[:MONITORED_BY]-
%(n:WaterStretch)
%RETURN DISTINCT qs as stations
% \end{lstlisting}

The query runtime was tested on a drainage network with 73 nodes and 73 edges, as well as three quality stations linked to the drainage network, with the longest path between water quality stations being 17 edges long. The query returned the result in 4ms.
As previously stated, path finding queries are simple to write in Cypher, and Neo4j, as a native graph database, enables the use of the benefits of graph theory rather than abstracting the information using tables.

% 4ms 
% 17 long
% 73nodes
% 72edges

\subsubsection{Q4: Finding quality stations contained in the same catchment area as a given land use} % % get water quality stations that share tehsame watershed as a type of landuse

The query returns all existing quality stations in the same watershed as the land-use area. The query first locates the watersheds containing the land use, and then collects all quality stations that monitored the drainage network within the watershed. To obtain the same information using a SQL implementation, multiple joins (e.g., at least three joins) would be required. The query was tested using the same use-case as the previous query, with a drainage network of 73 nodes and 73 edges, as well as three quality stations connected to the drainage network. The query took 5ms to return the result to the console.

%\begin{lstlisting}[caption={All quality stations in the same watershed as the land-use area},frame=tlrb, label={lst:watershedStations},language=SQL]{Name}
%MATCH (:LandUse{id: $id})-[:WITHIN]-
%>(ws:Watershed)
%WITH DISTINCT ws
%MATCH (qs:QualityStation)<
%-[:MONITORED_BY]-
%(:WaterStretch)-[:WITHIN]-
%>(ws)
%RETURN qs AS stations
%\end{lstlisting}

% 5ms
% 17 long
% 73nodes
% 72edges
\subsubsection*{Summary}

The queries implemented in the project for searching nodes and paths have execution times lower than 10ms. Table ~\ref{tab:queriestime} summarises the execution times of the queries explored in this section. 

 \begin{table}[htb]
\centering
\normalsize{\footnotesize
    \caption{Query execution times.}
    \label{tab:queriestime}
    \begin{tabular}{ | c | c |}
    \hline
     Query & Execution Time \\
 \hline \hline 
  Q1 & 3 ms\\ \hline
  Q2 & 6ms\\ \hline
  Q3 & 4ms\\ \hline
  Q4 & 5ms\\ \hline
    \end{tabular} }
\end{table} 

The Cypher turns queries simpler to construct because it allows to explicitly write the relationship patterns to find nodes or paths. Furthermore, the graph database supports direct data linking, allowing for quick data traversal by simply following the data connections. In comparison, relational database solutions would have multiple tables to store the same information, and querying would be more complex, requiring expensive JOIN operations or the use of multiple queries to achieve the same result.

\section{Usability}
%falar nos tempos de 5.4 como todas as quesries demoram menos de 10ms podemos concluir que a escolha da BD não afectará o desempenho da applicacão, visto os 100ms de tempo de espera estarem muito longe dos 10ms 
%ref: https://www.researchgate.net/publication/220893869_A_Study_on_Tolerable_Waiting_Time_How_Long_Are_Web_Users_Willing_to_Wait
The client application showed that it can render a large number of elements over the map without compromising the responsiveness and exploiting the resource consumption of the application. Moreover, the map rendering and generating charts with large datasets present response times  within the acceptable values for a good user experience (i.e, 100-200ms)\cite{Dabrowski2001}.

The queries used for gathering data from databases are simple and with average execution times less than 10ms, which is significantly lower than the 100ms, a common value for HTTP latency \cite{Bermbach2020}, and even lower than the 2s tolerable waiting time for information retrieval \cite{Nah2004}. Therefore, the choice of the database will not affect the performance of the client application.

\section{CONCLUSIONS}

This thesis introduces the development of an open-source framework capable of aggregating all types of information in a GIS environment, visualising and analysing large water quality-related datasets, and interactively exploring topological relationships between data. In order to solve the mentioned issues, this thesis proposes a graph data model to integrate and relate water quality between monitoring stations and land use, as well as to exploit the relationship between water quality in a watercourse and reservoirs associated with infrastructures. Instead of a relational database, a graph database (i.e Neo4j) is implemented and analysed in this thesis work. % falar aqui que se vai usar um abd de grafos em vez de uma bd relacional

%%%%%

The graph data model implemented in this thesis project allowed to connect all different information through the existing relationships between data, and it was demonstrated that it can natively store graph-like data, such as the water distribution network. Furthermore, the flexibility of the data model, as an inherent feature of a graph database, was useful for storing data with a variable number of attributes, such as water quality stations and data.

%%%%%
The development of the framework required choosing the technologies and design models that allow building a solution capable of scale, performance-oriented and able to run almost all machines, including with low computational power. The framework allows users to explore the relationships between data and enable other types of analysis supported by graph path-finding algorithms, such as the search for paths to the origin of contamination via specific chemical parameters or identification of the water sub-systems that will be affected by the contamination via network paths. 

%%% 

The query analysis tests showed noticeable performance for the path-finding queries. The Cypher (i.e., graph query language of Neo4j) simplifies query construction by allowing explicit writing of relationship patterns to find nodes or paths. Furthermore, the graph database supports direct data linking, which enables fast data traversal by simply following the data connections. In the map interface and data visualisation tests, the Framework demonstrated its ability to handle large collections of points while providing responsive map navigation to the user. Render performance was usually guaranteed when exploring a map with a large dataset, which can be attributed to good data flow design between web application components and HTML best practices. The plot also showed a responsive interaction during the test and was reasonably fast when plotting a large dataset, demonstrating good query and client application performance. However, the input file tests showed that upload time for larger water quality data was noticeable. The larger files contain numerous parameters that will be converted to attributes of the water quality data nodes. Furthermore, before sending the file to the database, the web server verifies the file structure and checks for invalid data. Therefore, the parsing time will increase with file size. The input file feature can be optimised at both the query and server implementation levels.

In conclusion, the graph data model and the developed framework are proofs of concepts that show promising results. Allows users to access data that was previously difficult to visualise and has proven to be advantageous in comparison to relational databases. The framework can be a useful tool for researchers to quickly monitor water pollutants circulation across the river network and water distribution systems.

\addtolength{\textheight}{-12cm}   % This command serves to balance the column lengths

%%%%%%%%%%%%%%%%%%%%%%%%%%%%%%%%%%%%%%%%%%%%%%%%%%%%%%%%%%%%%%%%%%%%%%%%%%%%%%%%

%%%%%%%%%%%%%%%%%%%%%%%%%%%%%%%%%%%%%%%%%%%%%%%%%%%%%%%%%%%%%%%%%%%%%%%%%%%%%%%%

%%%%%%%%%%%%%%%%%%%%%%%%%%%%%%%%%%%%%%%%%%%%%%%%%%%%%%%%%%%%%%%%%%%%%%%%%%%%%%%%

%%%%%%%%%%%%%%%%%%%%%%%%%%%%%%%%%%%%%%%%%%%%%%%%%%%%%%%%%%%%%%%%%%%%%%%%%%%%%%%%

\bibliographystyle{IEEEtran}

\bibliography{foo.bib}

\end{document}